\documentclass[twocolumn,pra,showpacs,superscriptaddress,floatfix]{revtex4}
\usepackage{mathptmx}
\usepackage[T1]{fontenc}
\usepackage[latin9]{inputenc}
\usepackage{verbatim}
\usepackage{amsmath}
\usepackage{graphicx}

\makeatletter
\@ifundefined{textcolor}{}
{%
 \definecolor{BLACK}{gray}{0}
 \definecolor{WHITE}{gray}{1}
 \definecolor{RED}{rgb}{1,0,0}
 \definecolor{GREEN}{rgb}{0,1,0}
 \definecolor{BLUE}{rgb}{0,0,1}
 \definecolor{CYAN}{cmyk}{1,0,0,0}
 \definecolor{MAGENTA}{cmyk}{0,1,0,0}
 \definecolor{YELLOW}{cmyk}{0,0,1,0}
 }

\makeatother

\begin{document}

\title{Perfect transmission and highly asymmetric light localization in
photonic multilayers}

\author{Sergei V. Zhukovsky}

\affiliation{Department of Physics and Institute for Optical Sciences, University
of Toronto, 60 St.~George Street, Toronto, Ontario M5S 1A7, Canada.}

\email{szhukov@physics.utoronto.ca}

\affiliation{Theoretical Nano-Photonics, Institute of High-Frequency and Communication
Technology, Faculty of Electrical, Information and Media Engineering,
University of Wuppertal, \\
Rainer-Gruenter-Str.~21, D-42119 Wuppertal, Germany.}

\begin{abstract}
General principles for the existence of perfect transmission resonances
in photonic multilayer structures are formulated in terms of light
interference described by recurrent Airy formulas. Mirror symmetry
in the multilayer is shown to be a sufficient but not necessary condition
for perfect transmission resonances. Asymmetric structures displaying
perfect transmission in accordance with the proposed principles are
demonstrated. A hybrid Fabry-Pérot/photonic-crystal structure of the
type $(\text{BA})^{k}(\text{AB})^{k}(\text{AABB})^{m}$ is proposed,
combining perfect transmission and highly asymmetric electric field
localization. Strength and asymmetry of localization can be controlled
independently, to be of use in tailoring non-reciprocal behavior of
nonlinear all-optical diodes.
\end{abstract}
\maketitle

\section{Introduction\label{sec:INTRODUCTION}}

Probably the simplest case of inhomogeneous media, photonic multilayers
are a good testing ground for structures with complex geometrical
properties such as aperiodic long-range order (see, e.g., \cite{MaciaReview}
and references therein). Indeed, the availability of simple, cheap,
and reliable computational methods often makes it possible to relate
geometrical and optical properties in an explicit manner. To name
a few examples, scaling and self-similar features in optical spectra
of quasiperiodic Fibonacci \cite{Kohmoto1,Kohmoto2,Macia1} and fractal
Cantor \cite{our2002,our2004,our2005} multilayers were recently found
to result from geometrical self-similarities of the underlying structure.
%
{}It is even possible to formulate general relations for spectral properties
of structures with arbitrary layer arrangement \cite{our2008}.

\begin{figure}[b]
\includegraphics[width=0.95\columnwidth]{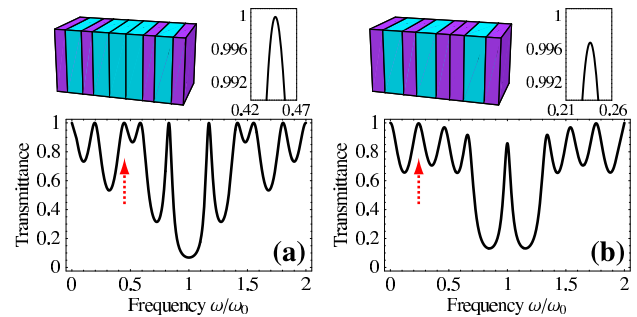}

\caption{(Color online) Example transmission spectra of (a) symmetric Cantor
multilayer $\text{BABAAABAB}$ and (b) non-symmetric Fibonacci multilayer
$\text{BABABBAB}$. $\text{A}$ and $\text{B}$ correspond to single
layers with $n_{\text{A}}=1.55$, $d_{\text{A}}=76\text{ nm}$ and
$n_{\text{B}}=2.3$, $d_{\text{B}}=113\text{ nm}$, so that $n_{\text{A}}d_{\text{A}}=n_{\text{B}}d_{\text{B}}=\pi c/2\omega_{0}=\lambda_{0}/4$
for $\lambda_{0}=700\text{ nm}$ as in \cite{FabioTacona09}. The
insets show an enlarged view of transmission peaks marked by arrows.\label{fig:intro}}

\end{figure}

One of rather intriguing properties of aperiodic multilayers is the
appearance of perfect transmission resonances (PTRs) in the optical
spectra, i.e., frequencies for which the multilayer has transmittance
exactly equal to unity ($|T|=1$). It is known that multilayers with
mirror symmetry (e.g., Cantor) commonly exhibit PTRs while those without
it (e.g., Fibonacci) usually do not: transmission peaks in such multilayers,
even if they look {}``perfect'', really have $|T|<1$ (see Fig.~\ref{fig:intro}).
Several accounts \cite{symHuang1,symHuang2,symHuang3gen,symMauriz4}
report PTRs if a Fibonacci structure is symmetrized and show that
perfect transmission is explicitly related to mirror symmetry \cite{symHuang3gen}.
However, more recent results show perfect transmission in asymmetric
multilayers based on periodic \cite{asymmHuang}, Fibonacci \cite{Nava09},
and Thue-Morse \cite{FabioTacona09} geometry. This suggests that
mirror symmetry is sufficient but not necessary for PTRs. 

Such PTRs in asymmetric structures are promising in designing non-reciprocal
optical devices such as nonlinear all-optical diodes \cite{FabioSPIE08}.
Indeed, an associated spatially asymmetric light localization at resonance
(see \cite{asymmHuang,FabioTacona09}) induces a non-reciprocal nonlinear
optical response, while perfect transmission assures that reflection
losses remain small. In this perspective, understanding the physical
principles of PTR formation in multilayers is undoubtedly of importance.
Most previous works, however, do not really arrive at such principles
beyond attributing PTR existence to {}``hidden symmetries'' in the
structures. Instead they draw rather formal conclusions in terms of
the widely employed transfer matrix method \cite{asymmHuang,Nava09}.
Such conclusions will benefit from an interpretation to reveal their
physical meaning. 

In this paper, the question of PTR presence in multilayer spectra
is addressed from another, more physical than computational standpoint.
Perfect transmission in any multilayer (however complex) is seen to
be governed by the same principles of multiple-beam interference as
in a simple Fabry-Pérot interferometer. Transmission and reflection
spectra of any multilayer are recovered using recurrent Airy formulas,
and conditions for any two structures to form PTRs when stacked together
are derived explicitly. From these conditions, known results such
as PTRs in mirror-symmetric multilayers naturally follow. Moreover,
it becomes possible to engineer structures with PTRs on purpose. As
an example, a structure comprising a Fabry-Pérot interferometer adjacent
to a 1D photonic crystal is proposed. This structure is shown to feature
both perfect transmission and highly asymmetric, strongly localized
electric field profile. Localization strength and asymmetry can be
controlled independently by structure design.

In Section~\ref{sec:AIRY}, theoretical background on using Airy-like
formulas for calculating the optical spectra of complex multilayers
is given. Section~\ref{sec:PRINCIPLES} follows with application
of these formulas to arrive at the principles encompassing all possible
cases of PTRs in multilayers. Specific cases such as mirror-symmetric
and Thue-Morse multilayers are considered, too. Section~\ref{sec:DESIGN}
further employs these principles proposing a design for a structure
featuring PTRs as well as strongly localized and highly asymmetric
electromagnetic field distribution. Finally, Section~\ref{sec:CONCLUSIONS}
summarizes the paper.

\global\long\def\sa{\text{\ensuremath{S_{1}}}}
\global\long\def\sb{\text{\ensuremath{S_{2}}}}
\global\long\def\sab{\text{\ensuremath{\bar{S}_{1}}}}
\global\long\def\sboth{\text{\text{\ensuremath{S_{1,2}}}}}

\section{Recurrent Airy formulas\label{sec:AIRY}}

\begin{figure}
\includegraphics[width=0.9\columnwidth]{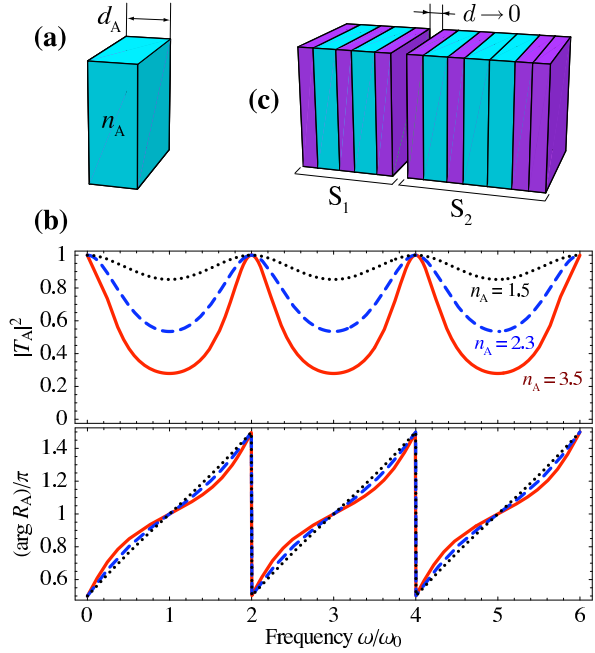}

\caption{(Color online) (a) A single layer and (b) its transmittance $|T_{\text{A}}|^{2}$
and phase shift of the reflected wave $\varphi_{\text{A}}$, given
by Eq.~\eqref{eq:Airy_layer}, for different values of $n_{\text{A}}$
{[}$d_{\text{A}}$ is chosen in accordance with Eq.~\eqref{eq:qwave}{]};
(c) an example of a composite $\sa\sb$ structure described by Eq.~\eqref{eq:Airy_stack}.\label{fig:illustration_theory}}

\end{figure}

We begin by considering a single dielectric layer (labelled $\text{A}$)
with refractive index $n=n_{\text{A}}$ and thickness~$d_{\text{A}}$,
located in a homogeneous dielectric medium with $n=n_{0}$ (Fig.~\ref{fig:illustration_theory}a).
Reflection and transmission coefficients of such a layer are given
by well-known Airy formulas (see, e.g., \cite{SunJaggard1,SunJaggard2})\begin{equation}
R_{\text{A}}=r_{0\text{A}}+\frac{t_{0\text{A}}r_{\text{A}0}t_{\text{A}0}e^{2i\delta_{\text{A}}}}{1-r_{\text{A}0}^{2}e^{2i\delta_{\text{A}}}},\quad T_{A}=\frac{t_{0\text{A}}t_{\text{A}0}e^{i\delta_{\text{A}}}}{1-r_{\text{A}0}^{2}e^{2i\delta_{\text{A}}}}.\label{eq:Airy_layer}\end{equation}
Here $\delta_{\text{A}}=(\omega/c)n_{\text{A}}d_{\text{A}}$ is the
phase accumulated by the wave in the layer. $r_{ij}$ and $t_{ij}$
are Fresnel reflection and transmission coefficients, respectively,
of an interface between the two media labelled by $i$~and~$j$
(the wave is incident on the interface from medium $i$~to medium~$j$).
Note that $R$~and~$T$ obtained by Eq.~\eqref{eq:Airy_layer}
are complex and contain information about amplitude as well as phase
of the reflected and transmitted wave. The {}``usual'' intensity-related
reflectance and transmittance are given by $|R_{\text{A}}|^{2}$ and
$|T_{\text{A}}|^{2}$, and it can be seen that $|R_{\text{A}}|^{2}+|T_{\text{A}}|^{2}=1$,
as is obvious from energy conservation. 

In such a simple system, the only frequency-dependent quantity is
the phase~$\delta_{\text{A}}$. Since $r_{0\text{A}}=-r_{\text{A}0}$
and $t_{0\text{A}}t_{\text{A}0}=1-r_{\text{A}0}^{2}$, it follows
that $T_{\text{A}}=1$ whenever $\delta_{\text{A}}=m\pi$ for integer~$m$.
Physically, this corresponds to constructive interference of forward-propagating
partial beams inside the layer, to occur when its optical thickness
is an integer multiple of a half-wave. Hence, a single layer features
equidistant PTRs like a Fabry-Pérot interferometer, albeit with poor-quality
mirrors (see Fig.~\ref{fig:illustration_theory}b). The PTR frequencies
are $2m\omega_{0}$ with $\omega_{0}$~defined by a well-known quarter-wave
(QW) condition \begin{equation}
(n_{\text{B}}d_{\text{B}}=)n_{\text{A}}d_{\text{A}}=\pi c/(2\omega_{0})=\lambda_{0}/4.\label{eq:qwave}\end{equation}

Similarly, let $\text{\ensuremath{\sa}}$ and $\sb$ be arbitrary
multilayers, e.g., arbitrary combinations of $\text{A}$~and~$\text{B}$
layers as in Fig.~\ref{fig:intro} (but of course, what follows remains
valid way beyond this example). Let the reflection and transmission
coefficients $R_{S}$ and $T_{S}$ be known for $S=\sa,\sb$, and
$\sab$ (a bar over $\sa$ denotes that $\sa$ is traversed in the
reverse direction). Inserting an infinitely thin layer of the ambient
medium between the structures (Fig.~\ref{eq:qwave}c), we can recover
the reflection and transmission for the composite $\sa\sb$ multilayer
stack:\begin{equation}
R_{\sa\sb}=R_{\sa}+\frac{T_{\sa}R_{\sb}T_{\sab}}{1-R_{\sab}R_{\sb}},\quad T_{\sa\sb}=\frac{T_{\sa}T_{\sb}}{1-R_{\sab}R_{\sb}}.\label{eq:Airy_stack}\end{equation}
Note that Eqs.~\eqref{eq:Airy_stack} follow from Eqs.~\eqref{eq:Airy_layer}
for $\delta=0$, and that the energy conservation $|T_{S}|^{2}+|R_{S}|^{2}=1$
holds. Also note that it is critical that both amplitude \emph{and}
phase of $R_{S}$ and $T_{S}$ is known. In a lossless, linear system
one can make use of time reversal to relate the spectra of $\sa$
and $\sab$ as $T_{\bar{S}}=T_{S}$, $R_{\bar{S}}/T_{\bar{S}}=-(R_{S}/T_{S})^{*}$.

By first taking $\sboth$ to be single layers with reflection and
transmission spectra given by Eqs.~\eqref{eq:Airy_layer} and then
using Eqs.~\eqref{eq:Airy_stack} and~\eqref{eq:Airy_layer} in
a recurrent fashion, we have a way to calculate transmission and reflection
spectra for a multilayer of any degree of complexity. Because such
recurrent calculation involves obtaining reflection and transmission
coefficients for many intermediate structures, it is numerically less
efficient than the transfer matrix method. However, the recurrent
procedure is often adopted for the sake of analytical insight into
the spectral properties of structures with internal symmetries, as
was demonstrated, e.g., for fractal multilayers \cite{SunJaggard1,SunJaggard2,our2004,our2005}.

%
{}

\section{Conditions for perfect transmission\label{sec:PRINCIPLES}}

Our goal is to formulate the existence conditions for a PTR in transmission
spectrum of an $\sa\sb$ stack. From Eq.~\eqref{eq:Airy_stack},
$|T_{\sa\sb}(\omega)|$ can be obtained as \begin{equation}
\left|T_{\sa\sb}\right|=\frac{\left|T_{\sa}\right|\left|T_{\sb}\right|}{\left|1-\left|R_{\sab}\right|\left|R_{\sb}\right|e^{i(\varphi_{\sab}+\varphi_{\sb})}\right|}\label{eq:PTR_cond}\end{equation}
where $\varphi_{\sab}$ and $\varphi_{\sb}$ are the phases of $R_{\sab}$
and $R_{\sb}$, respectively. Since $\left|R\right|^{2}+\left|T\right|^{2}=1$
in lossless structures, Eq.~\eqref{eq:PTR_cond} can be rewtritten
in the form\begin{equation}
\left|T_{\sa\sb}\right|^{2}=\frac{\left(1-\left|R_{1}\right|^{2}\right)\left(1-\left|R_{2}\right|^{2}\right)}{\left|1-\left|R_{1}\right|\left|R_{2}\right|e^{i\varphi}\right|^{2}},\label{eq:PTR_cond_2}\end{equation}
where we have denoted $\left|R_{1}\right|\equiv\left|R_{\sa}\right|=\left|R_{\sab}\right|$,
$\left|R_{2}\right|\equiv\left|R_{\sb}\right|$, and $\varphi\equiv\varphi_{\sab}+\varphi_{\sb}$
for brevity. If the denominator in Eq.~\eqref{eq:PTR_cond_2} is
non-zero, the PTR condition $\left|T_{\sa\sb}\right|=1$ is equivalent
to\[
\left[1-\left|R_{1}\right|^{2}\right]\left[1-\left|R_{2}\right|^{2}\right]=\left(1-\left|R_{1}\right|\left|R_{2}\right|\cos\varphi\right)^{2}+\left(\left|R_{1}\right|\left|R_{2}\right|\sin\varphi\right)^{2},\]
which reduces to\[
\left|R_{1}\right|^{2}+\left|R_{2}\right|^{2}=2\left|R_{1}\right|\left|R_{2}\right|\cos\varphi.\]
Obviously, this equation always holds if $\left|R_{1}\right|=\left|R_{2}\right|=0$,
which becomes one possible case for PTR, and never holds if $\left|R_{1}\right|=0$,
$\left|R_{2}\right|\neq0$ or vice versa. In all other cases $\left|R_{1}\right|\left|R_{2}\right|\neq0$
so we obtain\begin{equation}
\cos\varphi=\frac{\left|R_{1}\right|^{2}+\left|R_{2}\right|^{2}}{2\left|R_{1}\right|\left|R_{2}\right|}=1+\frac{\left(\left|R_{1}\right|-\left|R_{2}\right|\right)^{2}}{2\left|R_{1}\right|\left|R_{2}\right|}\geq1.\label{eq:cosine_condition}\end{equation}

If $\left|R_{1}\right|\neq\left|R_{2}\right|$, the right-hand side
of Eq.~\eqref{eq:cosine_condition} is strictly greater than unity,
so no PTR can exist because the condition $\cos\varphi>1$ cannot
be met. If $\left|R_{1}\right|=\left|R_{2}\right|$, PTRs can and
do occur whenever $\cos\varphi=1$.

For completeness, note that the limiting case when the denominator
in Eq.~\eqref{eq:PTR_cond_2} equals zero results in \[
1+\left(\left|R_{1}\right|\left|R_{2}\right|\right)^{2}-2\left|R_{1}\right|\left|R_{2}\right|\cos\varphi=0.\]
If $\left|R_{1}\right|\left|R_{2}\right|=0$, this equation is false.
Otherwise, it can be rewritten as \[
\cos\varphi=\frac{1+\left|R_{1}\right|^{2}\left|R_{2}\right|^{2}}{2\left|R_{1}\right|\left|R_{2}\right|}=1+\frac{\left(1-\left|R_{1}\right|\left|R_{2}\right|\right)^{2}}{2\left|R_{1}\right|\left|R_{2}\right|}\geq1,\]
and can only be satisfied if $\cos\varphi=1$ and $\left|R_{1}\right|\left|R_{2}\right|=1$.
Since the reflectance can never exceed unity, the latter implies that
$\left|R_{1}\right|=\left|R_{2}\right|=1$, i.e., the structure should
consist of two \emph{perfect} mirrors. Such an extreme case causes
the right-hand side of Eqs.~\eqref{eq:PTR_cond} and~\eqref{eq:PTR_cond_2}
to be indeterminate. This indicates that the approach based on the
interference of partial waves {[}Eq.~\eqref{eq:Airy_stack}{]} becomes
invalid with perfect mirrors when there are no partial waves to interfere.
However, this extreme can safely be ruled out by assuming that perfect
transmission is impossible in structures involving perfect mirrors.

As a result, we have obtained two possibilities for PTR existence.
First, when $\left|R_{1}\right|=\left|R_{2}\right|=0$, or, in the
original notation of Eq.~\eqref{eq:PTR_cond}, \begin{equation}
|T_{\sa}|=|T_{\sb}|=1.\label{eq:cond_1}\end{equation}
Second, when $\left|R_{1}\right|=\left|R_{2}\right|$ and $\cos\varphi=1$,
i.e., \begin{gather}
|T_{\sa}|=|T_{\sb}|\neq1,\label{eq:cond_2a}\\
\varphi_{\sab}+\varphi_{\sb}=2m\pi.\label{eq:cond_2b}\end{gather}

The first condition given by Eq.~\eqref{eq:cond_1} essentialy means
that whenever the individual structures $\sa$ and $\sb$ \emph{both}
have a PTR at \emph{exactly} the same frequency, the composite stack
$\sa\sb$ will always have a PTR at that frequency. In fact, this
conclusion could have been drawn directly from Eq.~\eqref{eq:PTR_cond}.
It is easily explained by the fact that if $\sa$ and $\sb$ are both
perfectly transparent, no reflection at the $\sa/\sb$ interface can
occur. Hence the incident wave is fully transmitted and there is no
possibility for the reflected wave to form. This is why, e.g., all
QW multilayers, where all layers conform to Eq.~\eqref{eq:qwave},
have PTRs at $\omega=2m\omega_{0}$ just as any one of the constituent
layers.

The second condition {[}Eqs.~\eqref{eq:cond_2a}--\eqref{eq:cond_2b}{]}
is more interesting because it explains how PTRs are formed in the
spectral regions of the composite structure where there were no PTRs
for either $\sa$ or $\sb$. Indeed, $\exp[i(\varphi_{\sab}+\varphi_{\sb})]=1$
in the denominator in Eq.~\eqref{eq:PTR_cond} renders it equal to
the numerator and causes $|T_{\sa\sb}|=1$ although $|T_{\sa}|=|T_{\sb}|\neq1$.
The PTR formation here can be explained by regarding the composite
structure as a Fabry-Pérot interferometer with very complex mirrors.
To begin with, the resonance occurs if the interference between partial
waves is constructive, i.e., if all the partial waves arising from
multiple reflection are in phase, as given by Eq.~\eqref{eq:cond_2b}.
Then, the resonance is perfect if the mirrors in the interferometer
are balanced and have equal reflectivity {[}Eq.~\eqref{eq:cond_2a}{]}.

\begin{figure*}
\includegraphics[width=1\textwidth]{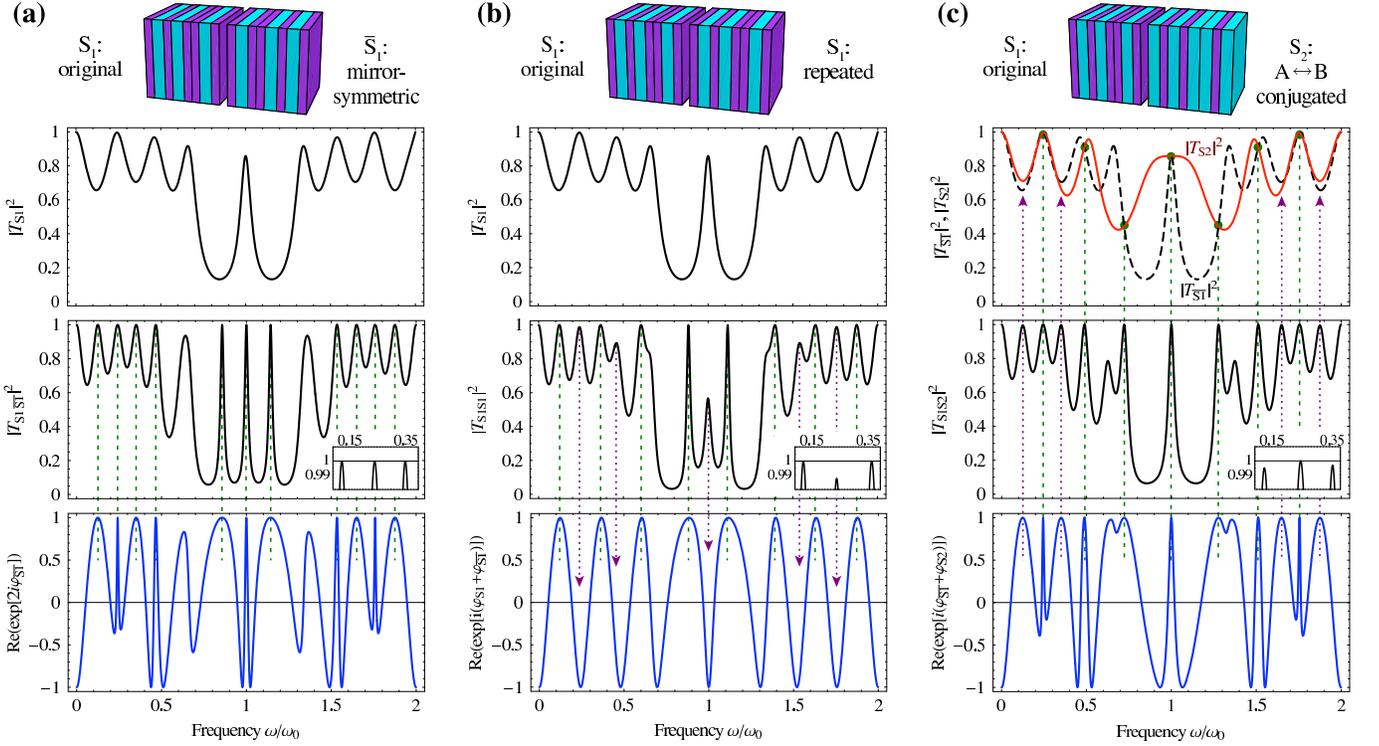}

\caption{(Color online) PTRs in composite structures $\sa\sb$: (a) mirror-symmetric
($\sb=\sab$); (b) double-stacked ($\sb=\sa)$; (c) Thue-Morse conjugated.
Top: transmission spectra of the constituent structures $\sboth$.
Middle: transmission spectrum of the whole structure. Bottom: spectral
dependence of the phase factor $\cos(\varphi_{\sab}+\varphi_{\sb})$
as in Eq.~\eqref{eq:cond_2b}. The dashed vertical lines show the
location of PTRs when both Eqs.~\eqref{eq:cond_2a}~and~\eqref{eq:cond_2b}
hold. The dotted lines with arrows designate the peaks that fail to
be PTRs due to violation of either Eq.~\eqref{eq:cond_2b} {[}in
(b){]} or Eq.~\eqref{eq:cond_2a} {[}in (c){]}. The insets represent
a blown-up view of some peaks to determine whether or not they are
PTRs. \label{fig:structure_examples}}

\end{figure*}

Eqs.~\eqref{eq:cond_2a}--\eqref{eq:cond_2b} let us easily see why
a mirror-symmetric structure readily supports PTRs while most other
structures do not. Mirror symmetry means $\sb=\sab$, so it is obvious
that $|T_{\sa}|=|T_{\sb}|$ and $\varphi_{\sab}=\varphi_{\sb}$ for
all frequencies. The only remaining condition to be fulfilled is Eq.~\eqref{eq:cond_2b},
i.e., $\varphi_{\sab}=m\pi$. Since the phase of the reflected wave
varies monotonically between transmission resonances in any multilayer
with rather few exceptions \cite{ourPhaseJOSA,ourPhasePRB}, there
should be numerous points where it crosses $m\pi$ (e.g., for one
layer it happens for $\omega=(2m-1)\omega_{0}$, see Fig.~\ref{fig:illustration_theory}b).
These points necessarily result in PTRs, as can be seen in Fig.~\ref{fig:structure_examples}a.
It is seen that for any PTR $\cos2\varphi_{\sab}=1$, except at $\omega=0$
and $\omega=2\omega_{0}$ where PTRs result from Eq.~\eqref{eq:cond_1}
rather than from Eqs.~\eqref{eq:cond_2a}--\eqref{eq:cond_2b}.

Another simple example would be $\sb=\sa$, i.e., when the same structure
is repeated twice in the stack. Again we have $|T_{\sb}|=|T_{\sab}|=|T_{\sa}|$
for all frequencies. However, Eq.~\eqref{eq:PTR_cond_2} here assumes
a different form, namely, $\varphi_{\sa}+\varphi_{\sab}=2m\pi$, which
is more difficult to satisfy (compare Figs.~\ref{fig:structure_examples}a
and~\ref{fig:structure_examples}b). As a result, the double-stack
structure $\sa\sa$ exhibits only half as many PTRs as does its mirror
symmetric counterpart $\sa\sab$. Both mirror symmetry and stack doubling
contribute to PTR formation in periodic structures, e.g., one-dimensional
photonic crystals. Note that if $\sa$~is asymmetric, so is $\sa\sa$,
and this case can be regarded as the simplest asymmetric multilayer
featuring PTRs.

Eqs.~\eqref{eq:cond_2a}--\eqref{eq:cond_2b} also encompass more
exotic cases involving intrinsically asymmetric structures. Consider
$\sa$ consisting of arbitrarily arranged $\text{A}$ and $\text{B}$
layers so that $n_{\text{A}}d_{\text{A}}=n_{\text{B}}d_{\text{B}}$
as in Eq.~\eqref{eq:qwave}, and $\sb$ obtained from $\sa$ by substitution
$\text{A}\leftrightarrow\text{B}$. The resulting structure is very
asymmetric (see Fig.~\ref{fig:structure_examples}c), yet it can
be shown to feature PTRs. This was observed by Nava et~al in \cite{Nava09}
for Fibonacci structures and further pointed out by Grigoriev and
Biancalana \cite{FabioTacona09} who named such structures {}``Thue-Morse
conjugated'' because one particular case of such structures, obtained
by repeatedly applying inflation rules $\text{A}\to\text{AB}$, $\text{B}\to\text{BA}$,
represents the well-known Thue-Morse sequence \cite{LiuThueMorse}.
Fig.~\ref{fig:structure_examples}c shows calculation results for
$\sa\sb$ with the same $\sa$ as for the previous examples (Figs.~\ref{fig:structure_examples}a--b).
Similarly to these, there are numerous frequencies where $\cos(\varphi_{\sab}+\varphi_{\sb})=1$
and Eq.~\eqref{eq:cond_2b} is satisfied, and each such frequency
represents a transmission peak. However, only a part of these peaks
turn out to be PTRs (see insets in Fig.~\ref{fig:structure_examples}),
namely the ones that simultaneously satisfy Eq.~\eqref{eq:cond_2a}.
The rigorous proof of how the fulfillment of these conditions results
from the Thue-Morse symmetry can be given and is expected to appear
in a forthcoming publication by Grigoriev et al. 

\begin{figure}
\includegraphics[width=0.6\columnwidth]{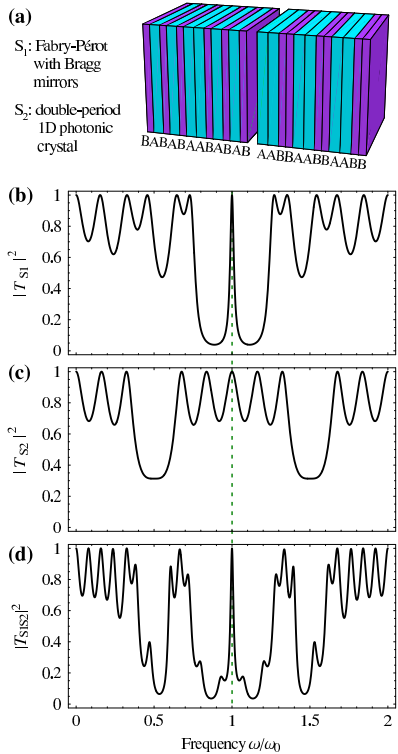}

\caption{(Color online) (a) The proposed design of a highly asymmetric multilayer
featuring PTRs given by Eq.~\eqref{eq:design}, along with transmission
spectra for (b) $\sa$, (c) $\sb$, and (d) $\sa\sb$. \label{fig:design_transmission}}

\end{figure}

\section{Perfect transmission \protect \\
in highly asymmetric structures\label{sec:DESIGN}}

Eqs.~\eqref{eq:cond_2a}--\eqref{eq:cond_2b} can be employed to
engineer a structure of any predefined geometry with a PTR at the
desired wavelength just by varying the refractive index and thickness
of the layers involved. Indeed, modifying $n_{\text{B}}/n_{\text{A}}$
in $\sb$ without violating Eq.~\eqref{eq:qwave} changes the value
of transmittance and reflectance while keeping the phases relatively
intact (see Fig.~\ref{fig:illustration_theory}b). This aids in fulfilling
Eqs.~\eqref{eq:cond_2a} and~\eqref{eq:cond_2b} simultaneously
and forms a PTR in the $\sa\sb$ structure. Subsequently varying~$\omega_{0}$
in Eq.~\eqref{eq:qwave} for both $\sa$~and~$\sb$ causes all
the spectra (both amplitude and phase) to scale uniformly, thus bringing
the PTR to the chosen value of the wavelength. Similarly designed
dual-interferometer structures of the type $(\text{AB})^{m}(\text{BA})^{m}(\text{A}'\text{B}')^{m}(\text{B}'\text{A}')^{m}$
were shown to possess PTRs \cite{asymmHuang}.

Our objective for this paper is to arrive at a design for multilayers
with highly asymmetric light localization at a PTR so as to facilitate
the non-reciprocal operation in a nonlinear optical diode \cite{FabioTacona09}.
A straightforward way to achieve the desired asymmetry is to stack
$\sa$ featuring a strongly localized mode with $\sb$ having an extended
mode, and to match the frequencies of the corresponding resonances.

An obvious choice for~$\sa$ with a maximally localized mode would
be a periodic QW multilayer with a half-wave defect, or, in other
words, a Fabry-Pérot interferometer surrounded by Bragg mirrors, so
that $\sa=(\text{BA})^{k}(\text{AB})^{k}$. If Eq.~\eqref{eq:qwave}
holds, a sharp transmission resonance occurs exactly at~$\omega_{0}$
(see Fig.~\ref{fig:design_transmission}b). On the contrary, the
modes are known to be maximally extended at $\omega=2\omega_{0}$
in any QW multilayer. By doubling the thickness of each layer, this
frequency can be halved to exactly match the resonance for~$\sa$.
A double-periodic photonic crystal structure of the type $(\text{AABB})^{m}$
can thus be used as $\sb$. The resulting stack the has the geometry
(Fig.~\ref{fig:design_transmission}a)\begin{equation}
\sa\sb=(\text{BA})^{k}(\text{AB})^{k}(\text{AABB})^{m}.\label{eq:design}\end{equation}

This design has the obvious advantage that both resonances in question
are PTRs (and they are exactly frequency matched), so there is no
need to go as far as Eqs.~\eqref{eq:cond_2a}--\eqref{eq:cond_2b}
and the resulting PTR in $\sa\sb$ is assured due to Eq.~\eqref{eq:cond_1}.
The second advantage is that frequency matching always occurs at $\lambda=\lambda_{0}/4$
so it is easy to design the structure for any desired wavelength using
any materials at hand. Fig.~\ref{fig:design_transmission}d confirms
the existence of a PTR, as does explicit numerical calculation of
$|T_{\sa\sb}(\omega_{0})|$, yielding 1 within limits of machine accuracy. 

\begin{figure}
\includegraphics[width=0.6\columnwidth]{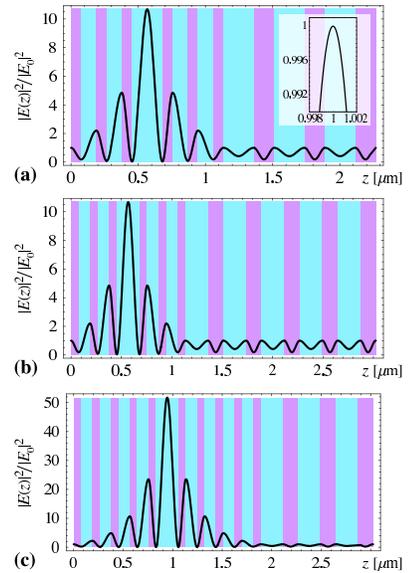}

\caption{(Color online) Electric field localization profile at the PTR frequency
$\omega=\omega_{0}$ (cf.~inset) for the proposed structure design
{[}Eq.~\eqref{eq:design}{]} (a) for $k=m=3$ as in Fig.~\ref{fig:design_transmission},
(b) for $k=3$, $m=5$ (enhanced asymmetry), and (c) for $k=5$, $m=3$
(enhanced localization strength).\label{fig:localization}}

\end{figure}

The electric field intensity distribution at the resonant frequency
is shown in Fig.~\ref{fig:localization}. It is seen that the localization
is clearly asymmetric and mainly present in~$\sa$. Using the same
materials as in Ref.~\cite{FabioTacona09}, comparable localization
strength is observed for a structure about 3 times thinner and having
32 layers instead of 64 (see Fig.~\ref{fig:localization}c). The
design of Eq.~\eqref{eq:design} also allows to control the localization
strength (by varying~$k$) and the asymmetry (by varying~$m$) independently
and in a wide range for a relatively minor change to the number of
layers (compare Fig.~\ref{fig:localization}a--c). This is opposed
to changing the number of generations in a Thue-Morse sequence, which
would double or halve the number of layers at once. A possibility
to build PTR-enabled structures with desired localization properties
using relatively few layers is important from the practical point
of view because losses would obviously be more detrimental to perfect
transmission in thicker structures \cite{Nava09}.

\begin{figure}
\includegraphics[width=0.9\columnwidth]{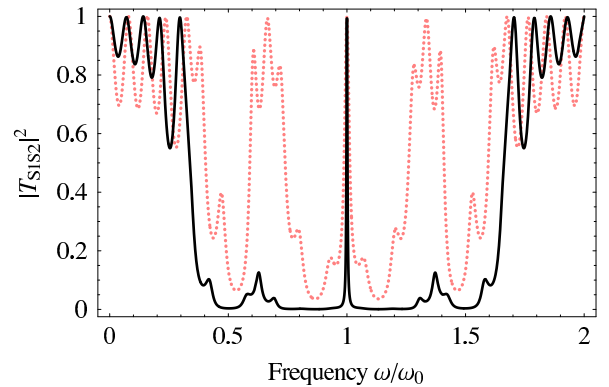}

\caption{(Color online) Transmission spectrum of the structure in Fig.~\ref{fig:design_transmission}a
for $n_{\text{A}}=1.55$, $n_{\text{B}}=2.3$ as in \cite{FabioTacona09}
(dotted) and for increased $n_{\text{B}}/n_{\text{A}}$ by setting
$n_{\text{A}}=1$ (solid).\label{fig:gap-overlap}}

\end{figure}
Note, finally, that the choice of geometry for $\sb$ is rather arbitrary
as any arrangement of $\text{AA}$ and $\text{BB}$ will produce the
same extended-mode PTR at $\omega_{0}$. This choice of geometry can
be regarded as an additional design tool to influence the transmission
spectrum around the PTR. For a periodic geometry $\sb=(\text{AABB})^{m}$
used in Eq.~\eqref{eq:design}, the two band gaps around $(1\pm1/2)\omega_{0}$
brought about by~$\sb$ (see Fig.~\ref{fig:design_transmission}c)
can overlap with the gap around~$\omega_{0}$ for~$\sa$ (Fig.~\ref{fig:design_transmission}b).
This would widen the region of predominantly low transmission surrounding
the designed PTR, which can prove useful. For the materials adopted
throughout this paper from Ref.~\cite{FabioTacona09}, it is not
yet the case, but the gap overlap can be achieved by increasing $n_{\text{B}}/n_{\text{A}}$
(Fig.~\ref{fig:gap-overlap}). It is seen that the PTR then becomes
very isolated in the transmission spectrum.

\section{Conclusions\label{sec:CONCLUSIONS}}

Using the formalism of recurrent Airy formulas, the conditions for
a mulilayer structure to exhibit perfect transmission resonances {[}Eqs.~\eqref{eq:cond_1}--\eqref{eq:cond_2b}{]}
are formulated rigorously in such a way that possibilities for PTRs
can be directly envisioned at the stage of structure design. Following
the previous results \cite{Nava09}, it was shown that mirror symmetry
is a sufficient but not necessary condition for PTR existence. PTRs
are shown to be possible in asymmetric structures, including Thue-Morse
conjugated multilayers \cite{FabioTacona09}. Based on frequency-matched
PTRs in structure parts, the design for a combined Fabry-Pérot/double-period
photonic crystal multilayer was proposed (Fig.~\ref{fig:design_transmission}a).
This structure was shown to feature perfect transmission resonances
with strongly localized and highly asymmetric spatial distribution
of electric field intensity (Fig.~\ref{fig:localization}). Strength
and asymmetry of localization can be controlled independently by changing
the design parameters, keeping the number of layers reasonably small.
It is expected that multilayers of this kind would enhance non-reciprocal
transmission if they contain nonlinear materials, improving the performance
of optical diodes and similar devices. 
\begin{acknowledgments}
The author is very grateful to Victor Grigoriev for inspiring and
fruitful discussions and helpful suggestions. Thanks go to Dmitry
Chigrin and all organizers of the international workshop on theoretical
and computational nanophotonics (TaCoNa-2009) for creating an excellent
environment for a very lively exchange of ideas. 

This work was supported in part by the Deutsche Forschungsgemeinschaft
(DFG Research Unit 557) and by the Natural Sciences and Engineering
Research Council of Canada (NSERC).\end{acknowledgments}

\end{document}